\documentclass[conference]{IEEEtran}
\IEEEoverridecommandlockouts
\usepackage{caption}
\usepackage{booktabs} 

\usepackage{cite}
\usepackage{amsmath,amssymb,amsfonts}
\usepackage{graphicx}
\usepackage{booktabs}
\usepackage{multirow}
\usepackage[table]{xcolor}
\usepackage{url}
\usepackage{textcomp}
\usepackage{algorithm}
\usepackage{algpseudocode}
\usepackage[utf8]{inputenc}
\usepackage{tcolorbox}
\tcbuselibrary{skins}
\DeclareUnicodeCharacter{FB01}{fi}

\def\BibTeX{{\rm B\kern-.05em{\sc i\kern-.025em b}\kern-.08em
    T\kern-.1667em\lower.7ex\hbox{E}\kern-.125emX}}
\begin{document}

\title{Pseudo2CodeQA: A Benchmark for LLM-Based Structured Algorithmic Reasoning in Code Generation}
\author{
  \IEEEauthorblockN{
    Shadikur Rahman\textsuperscript{*}\textsuperscript{†},
    Umme Ayman Koana\textsuperscript{*}\textsuperscript{†},
    Syed Muhammad Danish\textsuperscript{†}
  }
  \IEEEauthorblockA{\textsuperscript{*} York University, North York, Canada}
  \IEEEauthorblockA{\textsuperscript{†} Algoma University, Brampton, Canada}
  \IEEEauthorblockA{Emails: \{shadikur, ummekona\}@yorku.ca,\; syed.danish@algomau.ca}
}
\maketitle

\begin{abstract}

Large Language Models (LLMs) have achieved impressive performance in natural language-to-code generation; however, their ability to follow structured algorithmic reasoning remains insufficiently understood. We introduce \textbf{Pseudo2Code}, a benchmark designed to systematically evaluate the impact of structured pseudocode on code generation quality and algorithmic faithfulness. The benchmark consists of 300 manually validated real-world programming tasks spanning multiple domains and three difficulty levels (Easy, Medium, and Hard). Each task contains a problem description, structured pseudocode, reference implementation, and executable test suite. To ensure benchmark reliability, we adopt a dual-stage human validation protocol and release fully executable benchmark instances. Beyond the benchmark, we propose the \textbf{Pseudo2Code Agentic Framework}, a multi-stage pipeline that leverages pseudocode as an explicit intermediate reasoning representation for code generation. We evaluate both commercial and open-source language models using a rubric-based evaluation framework that measures correctness, completeness, relevance, clarity, reasoning quality, and pseudocode adherence, complemented by execution-based testing. Experimental results demonstrate that the proposed Pseudo2Code Agentic Pipeline consistently outperforms strong commercial and open-source baselines, achieving an overall score of 4.78 compared to 4.31 for the strongest baseline model. Furthermore, a human evaluation study involving 100 benchmark tasks shows strong agreement between human judgments and automated assessments. Our findings provide empirical evidence that structured pseudocode improves functional correctness, reasoning quality, and algorithmic faithfulness in code generation. We release Pseudo2Code to support future research on structured reasoning, interpretable code generation, and reliable AI-assisted software development.

\end{abstract}


\section{Introduction}

Large Language Models (LLMs) \cite{10.1145/3744255.3811741} have achieved remarkable progress in natural language--to--code generation \cite{rahman2026refactorcoderqa, 11391183}. Models such as Codex \cite{chen2021evaluating}, GPT-4 \cite{achiam2023gpt}, Code Llama \cite{roziere2023code}, and DeepSeek-Coder \cite{guo2024deepseek} demonstrate strong functional performance on execution-based benchmarks including HumanEval \cite{chen2021evaluating}, MBPP \cite{austin2021program}, and CodeXGLUE \cite{lu2021codexglue}. These benchmarks assess correctness by executing generated programs against unit tests, providing an objective measure of code synthesis quality. However, existing benchmarks predominantly evaluate code generation from unstructured natural language descriptions. The role of structured intermediate representations, particularly pseudocode, remains insufficiently understood. In classical algorithm design, pseudocode serves as an abstraction layer that decomposes algorithmic intent into structured procedural steps. It clarifies control flow, reduces ambiguity, and explicitly encodes logical dependencies. From a reasoning perspective, such structure should facilitate faithful translation into executable programs. Whether modern LLMs meaningfully leverage structured pseudocode or treat it as redundant textual input remains an open question. Prior research suggests that structured intermediate reasoning improves performance across domains. Chain-of-Thought (CoT) prompting \cite{wei2022chain} demonstrates that explicit reasoning traces enhance complex reasoning tasks. Program-of-Thought (PoT) prompting \cite{chen2022program} integrates executable program fragments into reasoning pipelines. Tool-augmented models \cite{schick2023toolformer, gao2023pal} and iterative self-refinement \cite{madaan2023self} further highlight the benefits of structured computational scaffolding. In code generation, execution-guided evaluation and repair mechanisms \cite{chen2021evaluating, jiang2023impact} have improved reliability. Nevertheless, there remains limited empirical evidence isolating the contribution of structured pseudocode as an explicit input representation for algorithmic code generation.

\begin{figure*}[t]
\centering
\includegraphics[width=\textwidth]{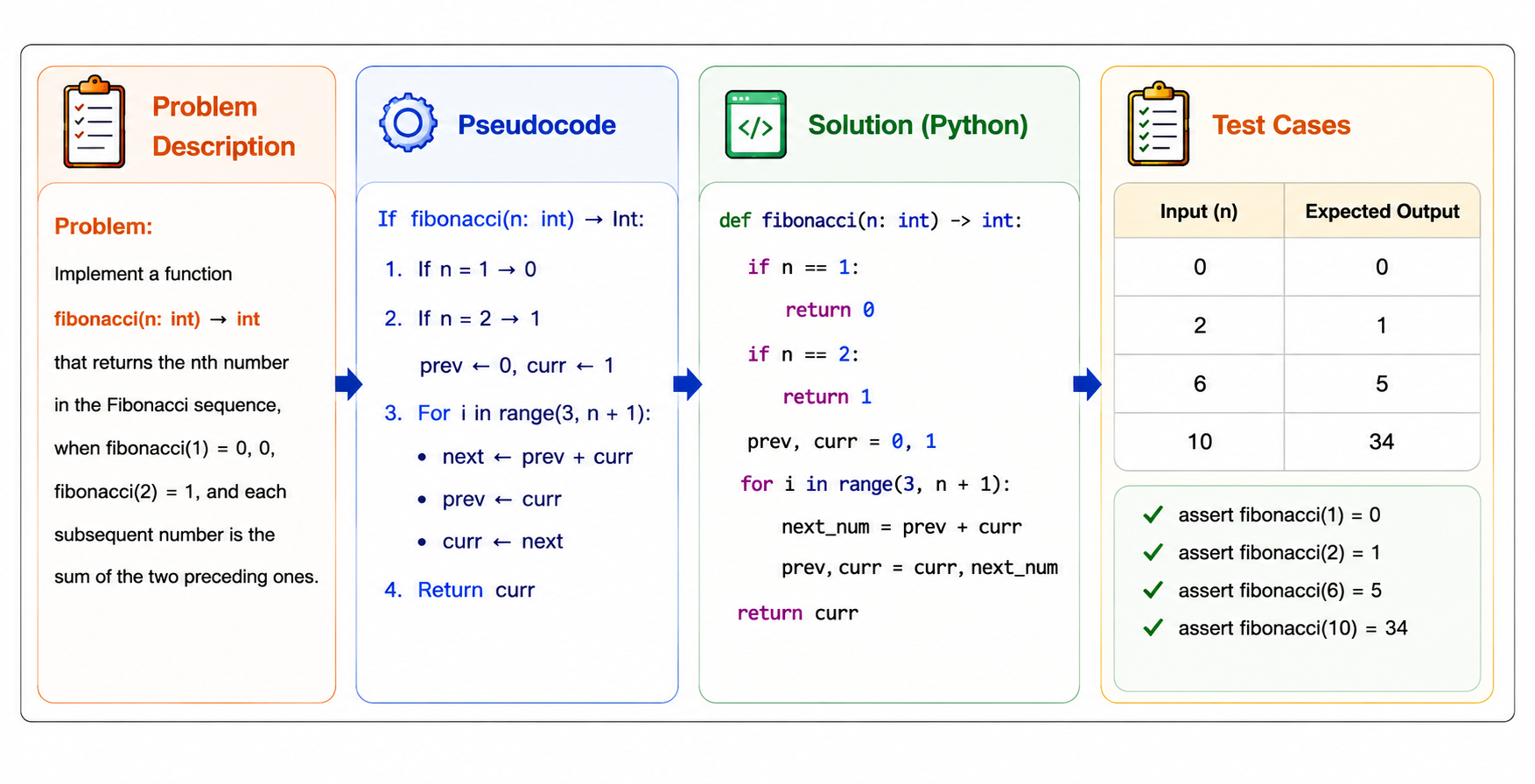}
\caption{Example benchmark instance in Pseudo2CodeQA. Each task consists of four aligned components: a natural-language problem description, a structured pseudocode representation, a reference Python implementation, and executable test cases. This design enables systematic evaluation of pseudocode-guided code generation and algorithmic faithfulness.}
\label{fig:pseudo2code_example}
\end{figure*}

To address this gap, we introduce \textbf{Pseudo2CodeQA}, a benchmark designed to systematically evaluate the impact of structured pseudocode on functional correctness and algorithmic faithfulness. The benchmark contains 300 real-world programming tasks collected from Stack Overflow, spanning data processing, financial computation, string manipulation, aggregation, and state tracking. Tasks are stratified into three difficulty levels (Easy, Medium, and Hard) to enable analysis across varying reasoning depth and control-flow complexity. Each task consists of four aligned components: (1) a formal problem description, (2) a structured pseudocode specification, (3) a reference Python implementation, and (4) comprehensive executable test cases. Figure~\ref{fig:pseudo2code_example} illustrates the standardized task format. To ensure reliability, we implement a dual-stage manual validation protocol in which independent human evaluators execute and verify all reference implementations against their test suites. Any discrepancies are corrected following controlled intervention rules, with transparent labeling of human-authored versus LLM-assisted modifications. The final release provides fully executable notebook files and verified test cases.

Pseudo2CodeQA supports three controlled input settings: (i) problem description only, (ii) pseudocode only, and (iii) combined problem plus pseudocode. This design enables direct quantification of the contribution of structured pseudocode as an intermediate reasoning representation. Beyond the benchmark, we propose the \textbf{Pseudo2Code Agentic Framework}, a multi-stage code generation pipeline built on top of \textbf{Gemini 2.5 Pro}. The framework explicitly incorporates pseudocode into the generation process through dedicated pseudocode generation, solution generation, test generation, and verification stages. By separating algorithmic planning from implementation, the framework enables systematic investigation of whether structured reasoning improves both functional correctness and algorithmic faithfulness compared to direct code generation approaches. To ensure a fair comparison, the direct generation baseline and the proposed agentic framework use the same underlying Gemini 2.5 Pro model, allowing improvements to be attributed primarily to the structured agentic workflow.

To evaluate benchmark performance, we assess both commercial and open-source language models using execution-based testing and a rubric-based evaluation framework that measures Correctness, Completeness, Relevance, Clarity, Reasoning, and Pseudocode Adherence. In addition, we conduct a human evaluation study on 100 benchmark tasks, where two independent evaluators assess 50 tasks each, to validate the reliability of the proposed evaluation framework and measure agreement between automated and human assessments. Our study is guided by the following research questions:

\begin{itemize}
\setlength\itemsep{2pt}
\item Does structured pseudocode improve functional correctness and algorithmic faithfulness in code generation?
\item How does pseudocode-guided code generation perform across varying levels of task complexity?
\item Can rubric-based automated evaluation reliably approximate human assessments of generated code quality?
\end{itemize}

Experimental results demonstrate that structured pseudocode substantially improves code generation quality. The proposed Pseudo2Code Agentic Framework achieves the strongest overall performance among all evaluated systems, obtaining an overall score of 4.78 and outperforming both commercial and open-source baselines. Furthermore, human evaluation results show strong agreement with automated assessments, providing evidence that structured pseudocode improves correctness, reasoning quality, and adherence to intended algorithmic workflows. This work makes the following contributions:

\begin{itemize}
\setlength\itemsep{2pt}

\item We introduce \textbf{Pseudo2CodeQA}, a benchmark consisting of 300 manually validated programming tasks for evaluating pseudocode-guided code generation and structured algorithmic reasoning. We make this dataset openly available at\footnote{\url{https://github.com/sadirahman/Pseudo2CodeQA}}.

\item We propose the \textbf{Pseudo2Code Agentic Framework}, a Gemini 2.5 Pro-based multi-stage code generation pipeline that explicitly incorporates pseudocode as an intermediate reasoning representation for improving code generation quality.

\item We conduct a comprehensive evaluation of commercial and open-source language models, providing empirical evidence on the effectiveness of structured pseudocode across varying task complexities.

\item We introduce and validate a rubric-based evaluation framework for measuring algorithmic faithfulness, reasoning quality, and implementation correctness beyond execution-based metrics.

\end{itemize}

By jointly providing a benchmark and an agentic framework, Pseudo2CodeQA establishes a reproducible foundation for studying structured reasoning, algorithmic faithfulness, and reliable code generation in large language models.

\section{Related Work}
\label{sec:rl}
\subsection{Code Generation Benchmarks}

Recent advances in large language models have significantly improved natural language–to–code generation. Benchmarks such as HumanEval \cite{chen2021evaluating}, MBPP \cite{austin2021program}, and CodeXGLUE \cite{lu2021codexglue} evaluate functional correctness by executing generated programs against unit tests. These datasets have become standard for assessing code synthesis quality and have enabled rapid progress in models such as Codex, GPT-4, Code Llama, and DeepSeek-Coder. More recent efforts have expanded evaluation to real-world software engineering tasks. SWE-bench \cite{jimenez2023swe} evaluates models on GitHub issue resolution, requiring multi-step reasoning and repository-level understanding. Similarly, datasets such as APPS \cite{hendrycks2021measuring} introduce more challenging competitive programming problems with complex reasoning requirements. While these benchmarks capture increasing task complexity, they primarily rely on natural language problem descriptions and do not explicitly incorporate structured intermediate representations such as pseudocode.

\subsection{Structured Reasoning in Language Models}

A growing body of work demonstrates that structured intermediate representations improve reasoning performance in LLMs. CoT prompting \cite{wei2022chain} shows that explicitly generating reasoning steps enhances performance on arithmetic and symbolic reasoning tasks. Extensions such as self-consistency \cite{wang2022self} further improve robustness by sampling multiple reasoning paths. PoT prompting \cite{chen2022program} introduces executable programs as intermediate reasoning steps, enabling models to offload computation to interpreters. Program-aided language models (PAL) \cite{gao2023pal} similarly integrate program execution into reasoning pipelines. Tool-augmented approaches such as Toolformer \cite{schick2023toolformer} demonstrate that models can learn to invoke external tools to improve reliability. While these approaches highlight the benefits of structured reasoning, they focus primarily on reasoning traces or executable programs rather than pseudocode as a human-interpretable intermediate representation. The role of pseudocode as a structured yet abstract representation for guiding code generation remains underexplored.

\subsection{Code Refinement and Self-Improvement}

Several works have investigated iterative refinement and self-correction in code generation. Self-Refine \cite{madaan2023self} enables models to iteratively improve outputs through self-feedback. Reflexion \cite{shinn2023reflexion} introduces a feedback loop where models critique and revise their own reasoning trajectories. Other approaches leverage execution feedback to repair incorrect programs \cite{chen2021evaluating, jiang2023impact}. These methods improve correctness by introducing feedback loops after generation. In contrast, our work focuses on improving generation quality \emph{before} execution by providing structured pseudocode as an explicit intermediate input.

\subsection{Intermediate Representations for Code Generation}

Intermediate representations have long been used in program synthesis and software engineering to bridge high-level specifications and executable code. In neural code generation, prior work has explored abstract syntax trees (ASTs) \cite{rabinovich2017abstract}, sketch-based generation \cite{dong2016language}, and modular decomposition \cite{yin2017syntactic}. These approaches enforce structural constraints during generation but often require specialized model architectures or supervision. More recent LLM-based approaches rely on prompting rather than architectural constraints, raising the question of how different input representations influence generation quality. While pseudocode is widely used in education and human problem-solving, its effectiveness as an input modality for LLM-based code generation has not been systematically evaluated.

In contrast to prior work, Pseudo2Code isolates structured pseudocode as an explicit intermediate representation and evaluates its impact under controlled conditions. By providing aligned problem descriptions, pseudocode, executable solutions, and validated test cases, our benchmark enables direct comparison between natural language and structured inputs. Furthermore, we introduce a rubric-based LLM-judge metric to assess algorithmic faithfulness beyond execution correctness, complementing existing evaluation methodologies.

\section{Benchmark Construction}

\begin{figure*}[t]
\centering
\includegraphics[width=\textwidth]{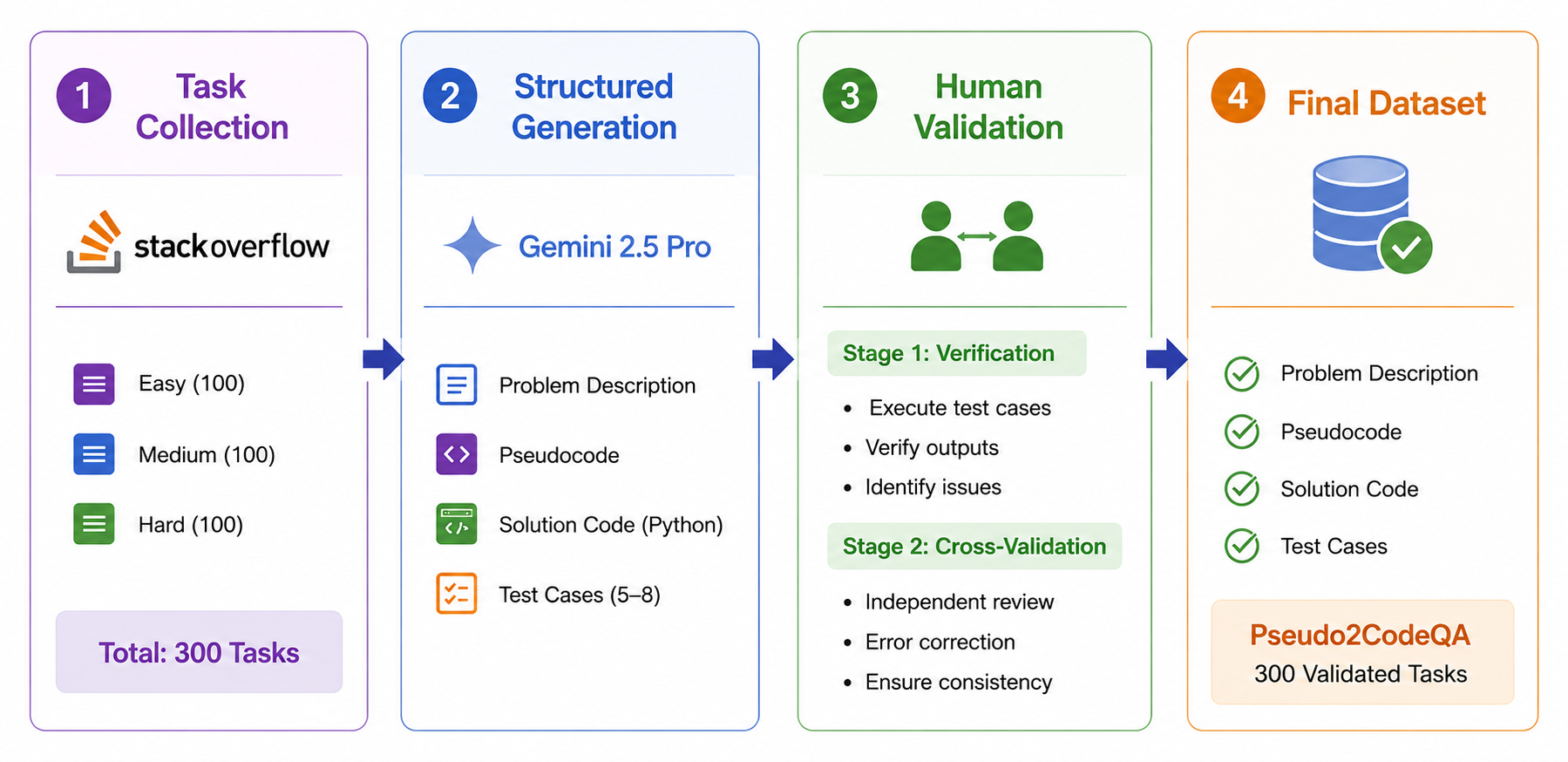}
\caption{Pseudo2CodeQA benchmark construction pipeline. Tasks are collected from Stack Overflow, transformed into aligned problem descriptions, pseudocode, solutions, and test cases using Gemini 2.5, validated through a dual-stage human verification process, and released as a clean benchmark with transparent modification labels for evaluating pseudocode-guided code generation.}
\label{fig:pipeline}
\end{figure*}

We present \textbf{Pseudo2CodeQA}, a benchmark designed to evaluate the role of structured pseudocode as an intermediate representation in code generation. An overview of the construction pipeline is shown in Figure~\ref{fig:pipeline}. The dataset is constructed through three stages: task collection, structured generation, and manual validation.

\subsection{Task Collection}

We construct a dataset of 300 programming tasks derived from real-world programming patterns observed in developer Q\&A platforms and software engineering practice. Instead of directly reusing existing problems, we abstract and reformulate recurring problem types into clean, executable tasks suitable for controlled evaluation. The dataset spans diverse real-world categories, including data processing, text manipulation, validation, aggregation, financial computation, scheduling, and workflow logic. To enable systematic analysis of reasoning complexity, tasks are stratified into three difficulty levels: \textit{Easy}, \textit{Medium}, and \textit{Hard}. \textbf{Easy} tasks involve single-pass logic and simple filtering or counting operations.  
\textbf{Medium} tasks require multi-step reasoning with conditional logic and state updates.  
\textbf{Hard} tasks involve multi-stage workflows, simulation, or complex rule-based reasoning. Each task is standardized into a concise and unambiguous problem description, ensuring consistency across the dataset.

\subsection{Structured Generation}

For each task, we construct aligned intermediate and executable representations using a controlled generation pipeline:

\begin{itemize}
    \setlength\parskip{0pt}
    \setlength\parsep{0pt}
    \setlength\itemsep{2pt}
    \item \textbf{Problem Description:} Generated using a structured prompt template to ensure clarity, determinism, and consistent input-output specification.
    
    \item \textbf{Pseudocode:} Produced using the LLM (Gemini 2.5 Pro) under strict formatting constraints. The pseudocode explicitly encodes algorithmic structure using standardized constructs (e.g., \texttt{FUNCTION}, \texttt{FOR}, \texttt{IF}, \texttt{RETURN}), ensuring consistency and interpretability.
    
    \item \textbf{Solution Code:} A reference Python implementation is generated to closely follow the pseudocode, preserving alignment between intermediate representation and executable logic.
    
    \item \textbf{Test Cases:} Each task includes 5–8 unit tests covering base cases, typical inputs, and edge conditions. These tests enable execution-based evaluation following prior work \cite{chen2021evaluating}.
\end{itemize}

This design ensures a one-to-one mapping between problem description, pseudocode, and solution code, enabling controlled comparison across input modalities.

\subsection{Manual Validation}

To ensure correctness, reliability, and reproducibility, we adopt a dual-stage manual validation protocol involving independent human evaluators and execution-based verification.

\paragraph{\textbf{Stage 1: Execution and Verification}}
Two independent evaluators execute each reference implementation against its corresponding test suite without modifying the code. The generated outputs are compared against the expected results to identify implementation errors, incorrect test cases, or inconsistencies between the problem specification and reference solution.

\paragraph{\textbf{Error Correction}}
When discrepancies are detected, corrections are applied following a controlled intervention protocol. Minor issues, such as syntax errors, incorrect variable references, or test-case inconsistencies, are corrected manually. More complex failures that require substantial algorithmic reasoning or code revision are resolved through LLM-assisted refinement using advanced code-generation models (e.g., GPT-4 or Gemini). All modifications are explicitly documented to ensure transparency and reproducibility.

\paragraph{\textbf{Stage 2: Cross-Validation}}
Following correction, a second evaluator independently reviews all modifications and re-executes the complete test suite to verify correctness and consistency. Based on the validation outcome, each task is assigned one of the following labels:

\begin{itemize}
\setlength\parskip{0pt}
\setlength\parsep{0pt}
\setlength\itemsep{2pt}
\item \textbf{Runnable}: The reference implementation passes all test cases without modification.
\item \textbf{LLM-Modified}: The implementation requires LLM-assisted refinement before passing validation.
\item \textbf{Human-Modified}: The implementation is corrected manually before passing validation.
\end{itemize}

This validation protocol ensures that all benchmark instances are fully executable, functionally correct, and reproducible while maintaining a transparent record of all modifications applied during dataset construction.

\subsection{Final Dataset}

The final dataset consists of 300 fully validated tasks, each containing a problem description, pseudocode, solution code, and verified test cases. All tasks are released as executable \texttt{.ipynb} notebooks to ensure reproducibility and ease of evaluation.

\begin{table}[h]
\centering
\small
\renewcommand{\arraystretch}{1.1}
\setlength{\tabcolsep}{10pt}

\begin{tabular}{lccc}
\toprule
\textbf{Difficulty} & \textbf{\#Tasks} & \textbf{Avg. LOC} & \textbf{Avg. \#Tests} \\
\midrule
Easy   & 100 & 8--12  & 3--5 \\
Medium & 100 & 15--25 & 4--6 \\
Hard   & 100 & 25--45 & 5--8 \\
\midrule
\textbf{Total} & \textbf{300} & \textbf{16--27} & \textbf{4--6} \\
\bottomrule
\end{tabular}

\caption{Dataset statistics for the Pseudo2Code benchmark. LOC denotes the number of lines of code in reference implementations.}
\label{tab:dataset}
\end{table}

This construction process results in a standardized and execution-ready benchmark that enables systematic evaluation of structured pseudocode in code generation.

\section{Pseudo2Code Agentic Framework}

\begin{figure*}[t]
\centering
\includegraphics[width=0.9\textwidth]{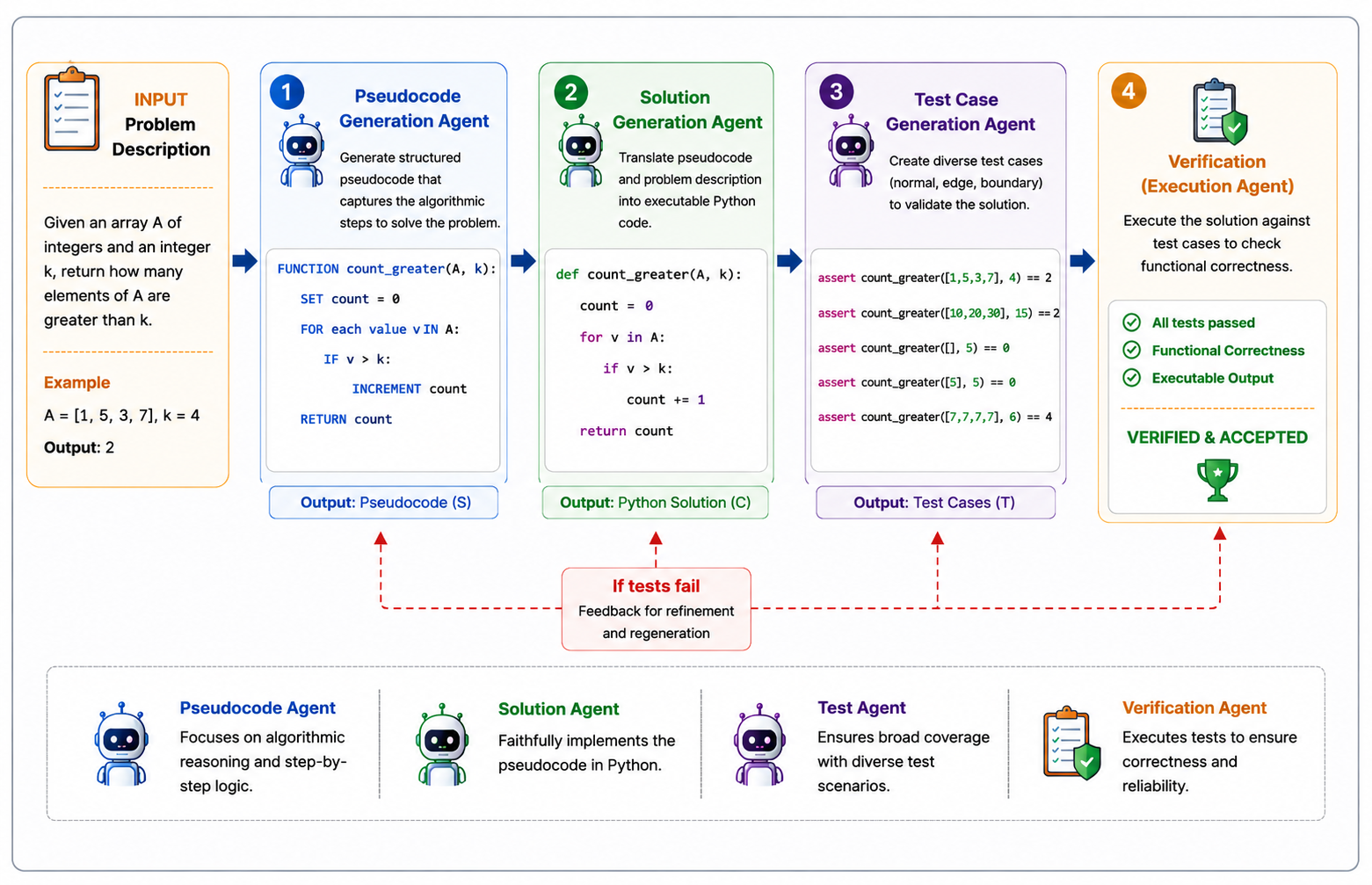}
\caption{The proposed Pseudo2Code Agentic Framework decomposes code generation into four specialized stages: pseudocode generation, solution generation, test case generation, and execution-based verification. By introducing pseudocode as an explicit intermediate reasoning representation, the framework separates algorithmic planning from implementation and validates generated programs through automated testing, improving transparency, faithfulness, and functional correctness.}
\label{fig:Pseudo2Code_Agentic1}
\end{figure*}

Beyond serving as a benchmark, Pseudo2Code enables the study of structured reasoning through an agentic code-generation framework. Traditional code generation systems typically translate a natural-language problem description directly into executable code using a single generation step. However, complex programming tasks often require intermediate reasoning to decompose the problem, identify algorithmic structure, and verify correctness. To address this limitation, we introduce the \textbf{Pseudo2Code Agentic Framework}, a multi-stage generation pipeline that explicitly incorporates pseudocode as an intermediate reasoning representation. As illustrated in Figure~\ref{fig:Pseudo2Code_Agentic1}, the framework decomposes code generation into four specialized agents: a \emph{Pseudocode Generation Agent}, a \emph{Solution Generation Agent}, a \emph{Test Generation Agent} and a \emph{execution-based verification}. This design encourages structured reasoning before implementation and provides an additional verification layer through automatically generated test cases.

\subsection{Pseudocode Generation Agent}

Given a programming task description $P$, the first agent generates structured pseudocode $S$ that captures the algorithmic intent of the problem. Rather than producing executable code directly, the agent focuses on identifying the core computational steps, control flow, variable updates, and decision logic required to solve the task. Formally,
\begin{equation}
S = f_{\text{pseudo}}(P)
\end{equation}

where $f_{\text{pseudo}}$ denotes the pseudocode generation model. The generated pseudocode serves as an interpretable reasoning trace that bridges the gap between natural-language requirements and executable implementations. By explicitly exposing intermediate reasoning, the framework reduces ambiguity and promotes algorithmic faithfulness.

\subsection{Solution Generation Agent}

The second agent receives both the problem description $P$ and the generated pseudocode $S$ and produces executable Python code $C$.

\begin{equation}
C = f_{\text{code}}(P,S).
\end{equation}

Unlike direct code generation approaches, the solution agent is constrained to follow the procedural structure encoded in the pseudocode. This encourages faithful translation of algorithmic reasoning into implementation details and reduces logical inconsistencies that commonly arise in single-step generation. The resulting implementation represents the candidate solution for the programming task.

\subsection{Test Generation Agent}

The third agent generates a set of executable unit tests $T$ for validating the generated solution $T = f_{\text{test}}(P,S,C)$. The generated tests include normal cases, boundary conditions, and edge cases designed to exercise different execution paths of the program. By leveraging information from both the pseudocode and generated implementation, the agent can create tests that better reflect the intended algorithmic behavior.

\subsection{Verification and Feedback}

The generated solution is executed against the automatically generated test suite. If all tests pass, the solution is accepted as correct. Otherwise, the failed test cases provide feedback indicating inconsistencies between the pseudocode, implementation, and expected behavior. This verification stage enables automatic assessment of functional correctness and provides a foundation for future iterative refinement or self-correction mechanisms.

\subsection{Advantages of Pseudocode-Guided Generation}

The proposed framework offers several advantages over direct natural-language-to-code generation. First, pseudocode provides an explicit intermediate reasoning layer that decomposes complex tasks into structured procedural steps. Second, the separation between reasoning and implementation improves transparency and interpretability. Third, automatically generated test cases provide an additional correctness signal that helps identify logical errors before deployment. Overall, the Pseudo2Code Agentic Framework investigates whether structured pseudocode can serve as an effective reasoning scaffold for code generation, improving functional correctness and algorithmic faithfulness compared to direct generation approaches. Table~\ref{tab:main_results} presents the overall performance of commercial models, open-source models, and the proposed Pseudo2Code framework on the 300-task benchmark.

\begin{table*}[t]
\centering
\caption{Performance comparison of commercial, open-source, and Pseudo2Code Agentic Pipeline on the Pseudo2CodeQA benchmark.
Scores are reported on a 1--5 scale and are assigned by GPT-5 using the proposed evaluation rubric.}
\label{tab:main_results}
\resizebox{\textwidth}{!}{
\begin{tabular}{lccccccc}
\toprule
\textbf{Model} &
\textbf{Correctness} &
\textbf{Completeness} &
\textbf{Relevance} &
\textbf{Clarity} &
\textbf{Reasoning} &
\textbf{Pseudocode Adherence} &
\textbf{Overall} \\
\midrule

GPT-4o Turbo & 4.23 & 4.21 & 4.20 & 4.22 & 4.16 & 4.19 & 4.20 \\
GPT-3.5 Turbo & 4.07 & 4.04 & 4.03 & 4.11 & 4.10 & 4.01 & 4.06 \\
Claude 3.5 Opus & 4.21 & 4.19 & 4.21 & 4.20 & 4.17 & 4.17 & 4.19 \\
Gemini 2.5 Pro & 4.33 & 4.32 & 4.32 & 4.33 & 4.27 & 4.29 & 4.31 \\

\midrule

DeepSeek-Coder-7B & 3.11 & 3.03 & 3.07 & 3.05 & 3.01 & 3.08 & 3.06 \\
DeepSeek-Coder-6.7B & 3.05 & 3.01 & 2.99 & 3.03 & 3.00 & 3.01 & 3.02 \\
DeepSeek-LLM-7B & 2.95 & 2.95 & 2.95 & 2.95 & 2.95 & 2.95 & 2.95 \\
DeepSeek-R1-Distill-Qwen-7B & 2.97 & 2.96 & 2.94 & 2.95 & 2.93 & 2.96 & 2.95 \\
DeepSeek-R1-Distill-Llama-8B & 2.99 & 2.94 & 2.98 & 2.92 & 2.97 & 2.99 & 2.97 \\
CodeLlama-7B-HF & 3.01 & 2.99 & 2.99 & 2.98 & 2.96 & 2.99 & 2.99 \\
Gemma-7B & 2.95 & 2.94 & 2.93 & 2.93 & 2.94 & 2.95 & 2.94 \\
Phi-2 & 2.97 & 2.95 & 2.95 & 2.93 & 2.93 & 2.94 & 2.95 \\
Mistral-7B & 2.95 & 2.96 & 2.94 & 2.95 & 2.92 & 2.93 & 2.94 \\

\midrule

DeepSeek-Coder-1.3B & 2.15 & 2.15 & 2.15 & 2.15 & 2.15 & 2.15 & 2.15 \\
DeepSeek-R1-Distill-Qwen-1.5B & 2.11 & 2.09 & 2.11 & 2.11 & 2.09 & 2.10 & 2.10 \\
LLaMA-3.2-3B & 2.45 & 2.50 & 2.44 & 2.41 & 2.43 & 2.41 & 2.44 \\
Gemma-2B & 2.19 & 2.17 & 2.15 & 2.16 & 2.15 & 2.15 & 2.16 \\

\midrule

Pseudo2Code (Direct) & 4.33 & 4.32 & 4.32 & 4.33 & 4.27 & 4.29 & 4.31 \\
\textbf{Pseudo2Code (Agentic Pipeline)} &
\textbf{4.75} &
\textbf{4.81} &
\textbf{4.79} &
\textbf{4.77} &
\textbf{4.78} &
\textbf{4.80} &
\textbf{4.78} \\

\bottomrule
\end{tabular}}
\end{table*}

\section{Results and Discussion}

In this section, we evaluate the effectiveness of the proposed \textbf{Pseudo2Code Agentic Pipeline} on the Pseudo2Code benchmark, which consists of 300 real-world programming tasks spanning three difficulty levels (Easy, Medium, and Hard). We first describe the evaluation metrics used to assess code generation quality, followed by a comparison against commercial and open-source language models. We then present a human evaluation study conducted on a subset of benchmark tasks and conclude with a discussion of the impact of structured pseudocode on algorithmic reasoning and code generation performance.

\subsection{Evaluation Metrics}

To comprehensively evaluate pseudocode-guided code generation, we employ a rubric-based evaluation framework that combines qualitative assessment with execution-based testing. Following the evaluation protocol, each generated solution is assessed across six dimensions: \textit{Correctness}, \textit{Completeness}, \textit{Relevance}, \textit{Clarity}, \textit{Reasoning}, and \textit{Pseudocode Adherence}. All qualitative metrics are scored on a 1--5 scale, where higher scores indicate better performance.

\textbf{Correctness} evaluates whether the generated solution produces the expected behavior and matches the intended functionality of the reference implementation. \textbf{Completeness} measures whether all required functionality and pseudocode steps are fully implemented. \textbf{Relevance} assesses whether the generated code remains focused on the specified programming task. \textbf{Clarity} measures code readability, organization, and maintainability. \textbf{Reasoning} evaluates the quality of the algorithmic approach and problem-solving strategy adopted by the model. Finally, \textbf{Pseudocode Adherence} measures how closely the generated implementation follows the provided pseudocode and intended algorithmic workflow.

\paragraph{LLM-as-a-Judge}
For rubric-based evaluation, we employ \textbf{GPT-5} as the evaluation model. Given the problem description, pseudocode, generated solution, reference implementation, and execution results, GPT-5 assigns scores for Correctness, Completeness, Relevance, Clarity, Reasoning, and Pseudocode Adherence according to the proposed evaluation rubric. In addition to qualitative assessment, we report the \textbf{Test Pass Rate}, defined as the percentage of unit tests successfully passed by a generated solution. The same evaluation protocol is adopted for human evaluation, enabling direct comparison between GPT-5-based and human assessments. The \textbf{Overall Score} jointly considers all qualitative dimensions with greater emphasis on Correctness, Pseudocode Adherence, and execution outcomes. To validate the automated evaluation framework, we conduct a human evaluation on 100 benchmark tasks, where two independent evaluators assess 50 tasks each using the same rubric.

\subsection{Main Results}

The results demonstrate that the proposed \textbf{Pseudo2Code Agentic Pipeline} achieves the strongest overall performance among all evaluated systems. The framework obtains an overall score of \textbf{4.78}, substantially outperforming the strongest commercial baseline, Gemini 2.5 Pro (4.31), and all open-source baselines. Across all evaluation dimensions, the proposed framework consistently achieves the highest scores, particularly in Completeness (4.81), Reasoning (4.78), and Pseudocode Adherence (4.80). Among commercial models, Gemini 2.5 Pro achieves the strongest baseline performance with an overall score of 4.31, followed by GPT-4o Turbo (4.20) and Claude 3.5 Opus (4.19). These models generate generally correct solutions but frequently deviate from the intended algorithmic workflow described by the pseudocode specification. Consequently, they achieve lower Reasoning and Pseudocode Adherence scores than the proposed framework. Open-source models exhibit substantially lower performance across all evaluation criteria. DeepSeek-Coder-7B achieves the strongest open-source performance with an overall score of 3.06, while most other models score below 3.0. These findings suggest that smaller open-source models often struggle to preserve algorithmic intent and accurately implement all required pseudocode steps.

An important observation is the performance gap between direct generation and the full agentic framework. While Pseudo2Code (Direct) achieves an overall score of 4.31, the Pseudo2Code Agentic Framework improves this to 4.78, representing a 10.9\% increase. This result indicates that the performance gains stem not only from the underlying foundation model but also from the proposed structured reasoning framework.

\subsection{Human Evaluation}

To validate the reliability of the GPT-5-based automated evaluation framework, we conducted a human evaluation study using 100 benchmark tasks. Two independent evaluators participated in the assessment, with each evaluator reviewing 50 tasks. The selected tasks were uniformly sampled across all difficulty levels to ensure balanced benchmark coverage.

\begin{table*}[t]
\centering
\caption{Human evaluation results on 100 benchmark tasks. Two independent evaluators assessed 50 tasks each using the proposed evaluation rubric. Scores are reported on a 1--5 scale.}
\label{tab:human_eval}
\resizebox{\textwidth}{!}{
\begin{tabular}{lccccccc}
\toprule
\textbf{Difficulty} &
\textbf{Correctness} &
\textbf{Completeness} &
\textbf{Relevance} &
\textbf{Clarity} &
\textbf{Reasoning} &
\textbf{Pseudocode Adherence} &
\textbf{Test Pass Rate} \\
\midrule

Easy
& 4.99
& 4.96
& 4.95
& 4.94
& 4.95
& 4.96
& 97\% \\

Medium
& 4.88
& 4.88
& 4.80
& 4.82
& 4.81
& 4.83
& 86\% \\

Hard
& 4.69
& 4.70
& 4.69
& 4.68
& 4.65
& 4.70
& 71\% \\

\midrule

\textbf{Overall}
& \textbf{4.85}
& \textbf{4.84}
& \textbf{4.81}
& \textbf{4.81}
& \textbf{4.80}
& \textbf{4.83}
& \textbf{84.6\%} \\

\bottomrule
\end{tabular}
}
\end{table*}

Table~\ref{tab:human_eval} summarizes the human evaluation results. Across all evaluation dimensions, annotators consistently assigned high scores, with Correctness achieving 4.85, Completeness 4.84, and Pseudocode Adherence 4.83. The overall execution pass rate reached 84.6\%, indicating that the majority of generated solutions successfully satisfied the provided test cases. The human evaluation further reveals the effect of task complexity. Easy tasks achieve near-perfect performance, obtaining a 97\% pass rate and scores approaching the maximum value across all evaluation dimensions. Medium tasks remain highly reliable, achieving an 86\% pass rate and average scores above 4.8. Hard tasks present the greatest challenge, resulting in a pass rate of 71\%; however, all qualitative metrics remain above 4.65, indicating that generated solutions generally preserve the intended reasoning process even when execution failures occur. Human annotators consistently reported that solutions generated by the proposed framework were easier to understand, more complete, and more faithful to the provided pseudocode than solutions generated by baseline models. These findings provide strong evidence that structured pseudocode improves both algorithmic faithfulness and implementation quality.

\section{Conclusion and Future Work}

In this paper, we introduced Pseudo2Code, a benchmark for evaluating the role of structured pseudocode in code generation. The benchmark contains 300 manually validated programming tasks spanning multiple domains and difficulty levels, each comprising a problem description, structured pseudocode, a reference implementation, and an executable test suite. We further proposed the Pseudo2Code Agentic Framework, which uses pseudocode as an intermediate reasoning representation to bridge natural language problem descriptions and executable code. Experimental results show that the proposed framework consistently outperforms commercial and open-source baselines in correctness, completeness, reasoning quality, and pseudocode adherence across both automated and human evaluations. Moreover, the strong agreement between human judgments and GPT-5-based evaluation demonstrates the reliability of the proposed assessment framework. Overall, the results show that structured pseudocode is an effective reasoning scaffold for improving code generation and aligning implementations with intended algorithms. Future work will extend the benchmark to additional programming languages, repository-level and multi-file code generation, and richer intermediate reasoning and execution-feedback mechanisms.

\bibliographystyle{IEEEtran}
\bibliography{pseudocode}
\end{document}